\documentclass[aps,prl,twocolumn,superscriptaddress,floatfix,amsmath]
{revtex4}

\usepackage{graphicx}

\newcommand{\dji}{\ensuremath{\Delta\lambda_{j;i}}} 
\newcommand{\exval}[1]{\langle #1\rangle} 

\newcommand{\pdji}{\ensuremath{P(\dji|\{\lambda\})}} 

\newcommand{\com}[1]{}

\begin{document}


\title{Mechanism for the failure of the Edwards hypothesis in the 
SK spin glass}
\author{P. R. Eastham}
\affiliation{Cavendish Laboratory, Madingley Road, Cambridge, CB3 0HE, UK}
\author{R. A. Blythe}
\affiliation{School of Physics, University of Edinburgh, Edinburgh EH9 3JZ, UK}
\author{A. J. Bray}
\author{M. A. Moore} 
\affiliation{School of Physics and Astronomy, University of Manchester, 
Manchester, M13 9PL, UK}

\date{\today}

\begin{abstract}
  The dynamics of the SK model at $T=0$ starting from random spin 
  configurations is considered. The metastable states reached by
  such dynamics are atypical of such states as a whole, in that the 
  probability density of site energies, $p(\lambda)$, is small at 
  $\lambda=0$.  
  Since virtually all metastable states have a much larger
  $p(0)$, this behavior demonstrates a qualitative failure of the Edwards
  hypothesis. We look for its origins by modelling the changes in the site
  energies during the dynamics as a Markov process. We show how the small
  $p(0)$ arises from features of the Markov process that have a clear physical
  basis in the spin-glass, and hence explain the failure of the Edwards
  hypothesis.
\end{abstract}


\maketitle

Complex systems like granular media have a large number of metastable
(blocked) configurations.  When shaken or tapped, they quickly relax
into another metastable state. A subsequent tap will result in another
blocked or jammed state, and so on.  The complexity (entropy) of
metastable states in granular systems or spin glasses is extensive in
the system size.  Edwards and co-workers have proposed that the
quasi-equilibrium steady state which results from repeated tapping can
be described using a thermodynamic measure over the metastable states
\cite{Edwards1994,EdwardsMehta}. The strongest version of such a
hypothesis predicts that a system adopts configurations which maximize
the entropy. In weaker versions parameters such as the energy or
volume are fixed, and the system adopts configurations which maximize
the entropy consistent with the constraints.

Edwards hypotheses have met with a high degree of success in many
complex systems. Some recent examples include predicting (i) the
distribution of contact forces \cite{metzger2005}, and the effective
temperature \cite{makse2002}, in simulations of granular media, (ii)
the dynamical entropy and correlation functions in the slow-dynamics
regime of the Kob-Anderson model \cite{barrat2000}, and (iii) the
distribution of steady-state energies in the tapped
Sherrington-Kirkpatrick model \cite{dean2003}. They seem to be a good
approximation, although not exact, for the zero-temperature
constrained dynamics of finite-dimensional Ising ferromagnets
\cite{godreche2005}. We note also support in the context of the slow
dynamics of mean-field spin-glass models, where it has been argued
that the effective temperature coincides with the Edwards temperature
\cite{godreche2005,barrat2000}. The underlying general idea that
dynamics does not strongly select amongst metastable states is yet
more widely used to attribute slow dynamics to a proliferation of
metastable states -- in optimization algorithms \cite{barthel2003},
for example.

Here we study dynamics in the the canonical SK model, for 
which the metastable states are already well-understood 
\cite{bray1980a,tanaka1980}. We show that the metastable states
selected by dynamics are of a very special character in which the
energy $2\lambda_i$ to flip the spin at site $i$ has a distribution
$p(\lambda)$ which is small for $\lambda \approx 0$.  Generic
metastable states have $p(0) \ne 0$.  The dynamically selected
metastable states are a vanishing fraction of the totality of
metastable states in the thermodynamic limit and therefore, according
to the Edwards hypothesis, should not be expected to be selected. We
provide a model of the dynamics which explains why it converges
onto this tiny subset of the metastable states.

The SK Hamiltonian is $H=-\sum_{(ij)} J_{ij} S_i S_j = 
-\frac{1}{2}\sum_i \lambda_i$,
where $S_i  = \pm 1$, $\lambda_i = S_i  \sum_{j\neq  i}  J_{ij}S_j$  
is  the  ``site-energy'',  equal  to one-half  of the energy  change on  
flipping the  spin $S_i$,  and $\Sigma_{(ij)}$ indicates a 
sum over  all pairs of  sites.  The interaction
strengths   $J_{ij}$   are  independent   random   variables  from   a
Gaussian distribution with zero mean  and standard deviation $1/\sqrt{N}$. 

We consider the non-equilibrium behavior of the model under single-spin 
relaxational dynamics 
\cite{glauberisingdynamics,coolen-glauberII,semerjian2004},
starting from a random initial state. We consider the $T=0$
limit of this dynamics, as in Refs.\
\onlinecite{dean2003,godreche2005} and \onlinecite{parisidynamics}, 
because it allows the
metastable states to be clearly identified. Further motivation for
studying this limit comes from its use in contexts ranging from
hysteresis in the SK model \cite{pazmandi1999} to domain growth in
ferromagnets; it corresponds to the basic Hopfield neural-network
algorithm, and to the greedy steps in the walk-SAT
algorithm \cite{barthel2003}.

The state evolves by flipping single spins with
$\lambda_i<0$, \textit{i.e.}, those which are opposed to the local
magnetic field on their site, until no such spins remain. Different
choices for the order of spin flips lead to different versions of the
algorithm. In the ``sequential'' algorithm a randomly selected
unstable spin is flipped at each timestep, while in the ``greedy''
algorithm the most unstable (minimum $\lambda_i$) spin is flipped.
The behavior of these different algorithms is remarkably
similar\cite{parisidynamics}.

\begin{figure}[t]
\includegraphics[width=246pt]{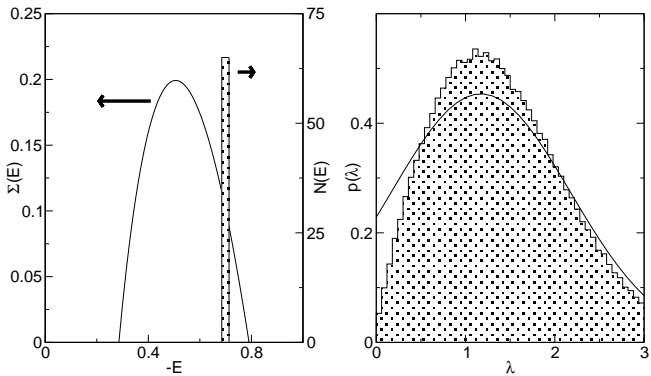}
\caption{\label{fig:sgresults} Discrepancies between simulations of
sequential  spin-glass dynamics  on a  system  of 5000  spins and  the
analytical predictions based  on flat-measure assumptions. Left panel,
curve  (left axis): Complexity of  metastable states  of  energy $E$,
$\Sigma(E)=\frac{1}{N}\log N_s(E)$ for the SK model, where $N_s(E)$ is
the  mean number  of metastable  states with  energy $E$.   Bar (right
axis):  Histogram  of the  converged  energies for  65  runs of  the
sequential   spin-glass  dynamics.    Right   panel,  curve:  Average
$p(\lambda)$ in  the metastable  states of energy  $-0.7$.  Histogram:
Average  $p(\lambda)$   over  the  final  states   of  the  spin-glass
dynamics.}\end{figure} 

The $T=0$ dynamics of the SK model converges onto
one-spin-flip-stable states, in which every spin aligns with its local
field.  This model is an attractive one in which to
consider the Edwards hypothesis, because these metastable states have
been studied analytically \cite{bray1980a,tanaka1980}. The key results
are shown in Fig.\ \ref{fig:sgresults}, in which the calculated and measured 
entropy (``complexity'') and distribution of local energies of the metastable 
states are compared. 
The converged energies do not cluster at the peak of the complexity 
curve, but are instead clustered in a narrow range around $E \approx -0.7$, 
so the dynamics certainly does not sample the metastable states uniformly.
Furthermore, the computed $p(\lambda)$ is qualitatively different from
the flat-average $p(\lambda)$ at the converged energy: The computed
$p(\lambda)$ has a negligible intercept, whereas the flat-average
$p(\lambda)$ has a significant finite intercept. Thus 
the dynamics does not uniformly sample the metastable states at the
converged energy. Furthermore, repeated ``tapping'' of a randomly
selected fraction of spins does not alter this conclusion: In our
simulations the steady-state does not develop an intercept. Thus the
states reached by the dynamics are always qualitatively different from
the totality of metastable states of the same energy, and in the
thermodynamic limit they are a negligible fraction of these states. In
other situations, it has been observed that the blocked states reached
by the dynamics have different energies to those typical of the
blocked states as a whole \cite{BergMehta}.  Our work shows this
feature too, but furthermore that the dynamically generated states are
even atypical of the states of the same energy.

To understand why the typical metastable states are not realized we
must look to the dynamics. We simplify the problem by considering
only the population of site-energies, $\{\lambda\}$, and making the
working assumption that the evolution of $p(\lambda)$ can be modeled
in terms of a Markov process in this population.

The population dynamics is designed to parallel the real spin-glass
dynamics. At each step an unstable spin $i$ is flipped, corresponding
to $\lambda_i \to -\lambda_i$. In the spin-glass the other site
energies $\lambda_j$ shift by an amount
\begin{equation} \Delta\lambda_{j;i} =  -2 S_i S_j J_{ij}.
\label{eq:driftdefn}\end{equation} 
Here $S_i$ and $S_j$ denote the spin configuration before the flip. To
obtain a population dynamics we replace the drifts $\dji$ with
functions of the site-energies. In the Markov approximation we
replace them with independent random variables, whose distribution
$\pdji$ depends only upon the site-energies at each step.

Similar approaches have previously been applied to the SK model
\cite{horner,coolen-glauberII}, granular media \cite{Edwards1994}, the
walk-SAT algorithm \cite{barthel2003}, and spin models on random
graphs \cite{semerjian2004,BergMehta}. Previous work on the SK model
has attempted to calculate $\pdji$.  Although this approach has met
with some success \cite{horner}, it leads to very involved models.
Owing to their complexity, these models are only tractable
numerically, and their physics remains obscure. We therefore take a
different approach, which is to determine the general features of
$\pdji$ that suffice for a qualitative understanding of the dynamics.

We can deduce some of the general features of $\pdji$ directly from
(\ref{eq:driftdefn}). Because the model is completely connected, summing the
drifts over all the unflipped spins gives the sum rule
$\sum_{j\neq i}\Delta\lambda_{j;i} = -2\lambda_i$.
Therefore, to model the dynamics with a Markov process, we must take 
$\pdji$ to have a mean $\propto 1/N$ in the large-$N$ limit.
Since $S^2=1$, the variance of the drifts is then just associated with
that of the bond distribution,
\begin{equation}
\exval{\dji^2}-\exval{\dji}^2 \sim 4/N. 
\label{eq:variance} \end{equation} 

\begin{figure}[b]
\includegraphics[width=246pt]{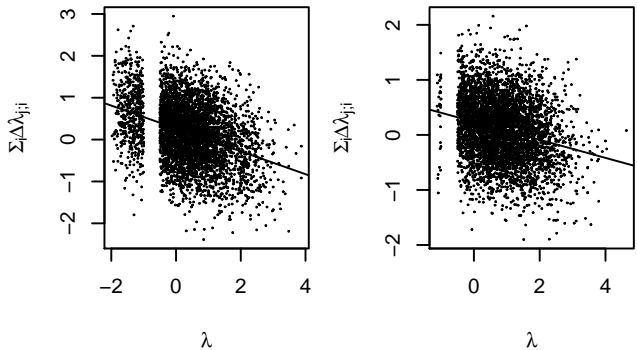}
\caption{\label{fig:driftnflips} Total changes in the site energies, 
$\sum_{i}\Delta \lambda_{j;i}$ of unflipped  spins when all spins with
site  energies in  the  ranges $-1.0<\lambda_i<-0.5$  are flipped,  in
configurations generated  by 100 (left panel, 671 flipped spins), and
500 (right panel, 420 flipped  spins) steps of the greedy algorithm on
a system  of 5000 spins.   This algorithm converged after  2465 flips.
The straight  lines show  linear fits to  the data. 
}
\end{figure}

Our simulations of the spin-glass dynamics converged in $\lesssim N$
flips. Since $J_{ij} = O(1/\sqrt{N})$, the third and higher
cumulants of $\dji$ are higher order in $1/N$ than the mean and
variance. Therefore the total drift produced by the higher cumulants
is negligible over the convergence time, and we may take $\pdji$ to be
Gaussian. Any correlations between the drifts, $\dji$, and the fields
$\lambda_j$ would have a qualitative effect on the evolution of
$p(\lambda)$. We looked for such correlations by taking the states
generated during the spin-glass dynamics and numerically evaluating
the drifts when spins are flipped.  The results are shown in Fig.\ 
\ref{fig:driftnflips}. Each point is the total drift of an unflipped
spin  as   a  function  of   its  site-energy  when  all   spins  with
site-energies in a small range are flipped.

Note the general correlation between the drifts and the site-energies
which can be seen in Fig.\ \ref{fig:driftnflips}. The overall drift on
flipping a spin $i$ is fixed by the sum rule 
but it is non-uniformly distributed amongst spins according to their
energies: Highly unstable spins tend to have their site-energies
strongly increased, at the expense of a reduced increase or a decrease
in the site-energies of the more stable spins.  This is physically
reasonable because a very unstable spin has mostly unsatisfied bonds,
while a very stable spin has mostly satisfied bonds. Therefore the
spin $i$ is likely to be connected to a highly unstable spin by an
unsatisfied bond, and to a highly stable spin by a satisfied bond,
producing the observed correlation.

We now consider whether the general features we have identified can
explain aspects of the spin-glass dynamics, in particular the
observation that it apparently converges in $\sim N$ timesteps, to a
state with a small intercept and an approximately linear $p(\lambda)$.
We adopt the following minimal model, which captures the behavior of
the distribution $p(\lambda)$ at small $\lambda$ and at late times. We
make the simplest assumption, that the drift $\dji$ in the value of
$\lambda_j$ resulting from flipping an unstable spin $i$ is a Gaussian
random variable with mean $c/N$ ($c>0$) and variance $\sigma^2/N$,
where, according to Eq.\ (\ref{eq:variance}), $\sigma^2 = 4$. This
assumption is motivated by the correlations visible in Fig.\
\ref{fig:driftnflips}, which lead us to expect that the mean drift of
a low-energy spin is non-vanishing as the converged state is
approached. Since the assumption of a constant drift violates the
previously derived sum-rule, it cannot be correct for {\em all} sites.
Our model is designed to address the behaviour of $p(\lambda,t)$ at
small $\lambda$.

The equation of motion for $p(\lambda,t)$ is, for
large $N$,
\begin{eqnarray}\label{eq:seqpopeqn} 
\frac{\partial p(\lambda,t)}{\partial t} 
= \frac{1}{q(t)}\left[p(-\lambda,t)\theta(\lambda)-p(\lambda,t)
\theta(-\lambda)\right] \nonumber \\ 
-c\,\frac{\partial p(\lambda,t)}{\partial\lambda}
 + \frac{\sigma^2}{2}\,\frac{\partial^2 p(\lambda,t)}{\partial\lambda^2},
\end{eqnarray} 
where $q(t)=\int_{-\infty}^0 p(\lambda,t) d\lambda$ is the weight in the 
negative side of the distribution (from which the flipped spins are drawn)
at time $t$, and the units of time are such that there are $N$ moves per 
unit time. 

The  first  term in  (\ref{eq:seqpopeqn})  derives  from the  flipping
process  $\lambda_i  \to   -\lambda_i$,  which  simply  transfers  the
population from negative to positive $\lambda$ at a rate of 1 spin per
timestep. The second  term derives from the mean  of the drifts, which
leads,  within our  model, to  a uniform  convection in  the $\lambda$
space.  The  final diffusion  term is due  to the fluctuations  in the
drifts. All these processes occur  on the same timescale, taking $\sim
N$ steps,  or a time $\sim  N^0$, to produce  an effect of order  1 on
$p(\lambda,t)$.

\begin{figure}[b]
  \includegraphics[width=242pt]{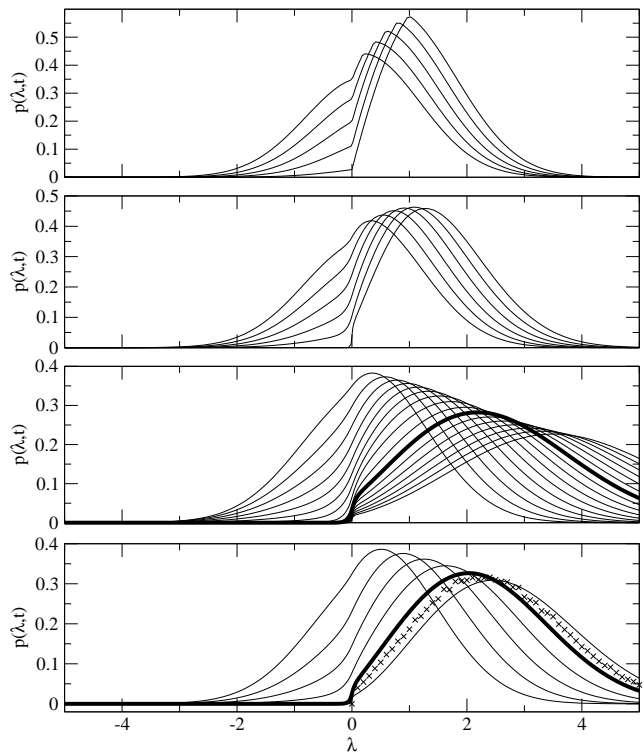}
\caption{\label{fig:numericalsol} The solution to (\ref{eq:seqpopeqn}), 
  with a Gaussian initial condition and $c=4$ (top three panels) and $8$ 
  (lowest panel), for $\sigma^2=0$ (top panel), $\sigma^2 = 1$(middle panel), 
  and $\sigma^2 = 4$ (lower two panels). Curves are plotted at time 
  intervals of 0.05. The bold curves in the lower two figures are the
  earliest at which $p(0,t)\approx 0.03$. They agree with the histograms
  obtained by direct simulation of the population dynamics model with $5000$
  fields, shown for $c=8$(crosses).}
\end{figure}

To understand the solutions to Eq.\ (\ref{eq:seqpopeqn}) we first
consider the case $\sigma^2=0$.  The equation of motion can then be
solved analytically, to give $p(\lambda,t)$ in terms of integrals over
$p(\lambda,0)$.  The results are shown in the top panel of Fig.\
\ref{fig:numericalsol}, for a Gaussian initial condition and
$c=4$. The number of spins with $\lambda<0$ is always decreasing at a
finite rate, due to the convection across $\lambda=0$ and the flipping
process. Thus this process certainly converges, reaching $q(t)=0$ in a
finite time. In general the decay of $p(0,t)$ near the end of the
evolution is linear in time, which combines with the convection to
produce a linear $p(\lambda)$, with no intercept, in the converged
state.  The slope depends on the initial conditions and on $c$. It
diverges as $c\to 0$, where the resulting $p(\lambda)$ is just the
half-Gaussian created by the flipping.

For $\sigma^2>0$, we have solved (\ref{eq:seqpopeqn}) numerically. The
resulting $p(\lambda,t)$ are shown in the lower three panels of Fig.\ 
\ref{fig:numericalsol}.  For these values of $c$ and $\sigma^2$ the
behavior at early times is similar to that with $\sigma^2=0$. The
diffusion, however, smooths out the singularities (discontinuity of
slope at $\lambda=0$) evident in the $\sigma^2=0$ solutions, and
broadens the distribution, but the tail of unstable spins continues to
decay at a significant rate. This can be understood by noting that
while the positive slope at $\lambda=0$ leads to a diffusion current
back towards $\lambda<0$, for these parameters this current is too
small to overcome the loss due to flipping and convection.  In
contrast, if $c$ is too small the solution with $\sigma=0$ would have
a large average slope at $\lambda=0$, and the diffusion would have a
major effect.

Although for some $c$ and $\sigma^2$ the early-time behavior is similar to
that of the model with $\sigma^2=0$, we see that a new regime appears at later
times. As the tail of unstable spins becomes narrower, the slope at
$\lambda=0$ increases, while the intercept continues to decay. This slows the
decay of $q(t)$, which obeys
$dq/dt =-1-cp(0,t)+ (\sigma^2/2)\,(\partial p/\partial\lambda)|_{\lambda=0}$,
with terms due to flipping, convection, and diffusion respectively.
Indeed, in the the lower two panels the slope at $\lambda=0$ is approaching
the critical slope of $2/\sigma^2$ at which the diffusion current balances the
loss due to flipping.  $q(t)$ must continue to decay, since the bulk of
$p(\lambda)$ will continue to diffuse and convect, and by continuity this must
reduce the tail of unstable spins. However this decay is extremely slow.
Furthermore, it is an artifact of our low-energy approximation, in which we
replaced the $\lambda$-dependent convection rate by a constant. In a more
complete treatment the biasing visible in Fig.\ \ref{fig:driftnflips} would
tend to confine the bulk of the distribution to a region centered on $\lambda
\sim 1$, due to negative convection rates at large $\lambda$, whereas in the
model the maximum of the distribution continues to drift to the right -- see
Fig.\ \ref{fig:numericalsol}.

In an infinite  system the Markov process and  the spin-glass dynamics
terminate when $q(t)=0$. Our numerics suggest that this does not occur
in a finite  time for the Markov process,  unless $\sigma=0$. Hence it
is inconsistent with the conjecture  that the dynamics of the infinite
spin-glass converges in a finite time. However, for moderate values of
$c$ the  features in $p(\lambda)$  associated with the slowing  of the
dynamics become so small that it would require a very large system for
them  to be  resolved. Therefore  we suggest  that the  minimal Markov
model  may be  adequate  to  understand the  convergence  seen in  the
spin-glass simulations, which are finite, albeit large.

In the finite spin-glass the converged $p(\lambda)$ has a small
intercept, which we can estimate by fitting to histograms such as
those shown in Fig.\ \ref{fig:sgresults}.  For $N=1000$ we obtain an
intercept of $0.06$, and $\approx 0.03$ for $N=5000$ and $N=10000$,
consistent with the intercept of $2/\sqrt{N}$ suggested in Ref.\
\onlinecite{parisidynamics}. This scaling is explained by the Markov
model, since for $p(0)\lesssim 1/\sqrt{N}$ the average diffusion flux
from positive to negative $\lambda$ is less than the one spin per
timestep transferred in the opposite direction by the flipping. The
dynamics will rapidly converge after such an intercept is reached,
with little further change in $p(\lambda)$.

Based on these arguments and the results for the direct simulations of
the spin glass, we suggest that the Markov process will converge in a
finite system when $p(0,t)$ obtained from Eq.\ (\ref{eq:seqpopeqn})
becomes comparable with $1/\sqrt{N}$. For a large enough $c$, this
condition is met before the dynamics becomes dominated by diffusion,
and the resulting $p(\lambda,t)$ has some features similar to that of
the simulational result. This can be seen in the lower two panels of
Fig.\ \ref{fig:numericalsol}, where we mark in bold the $p(\lambda,t)$
at which $p(0,t)\approx 0.03$. This corresponds to the smallest
intercept we have seen in the spin-glass simulations. Direct
simulations of the minimal model in a finite population lead to
similar distributions.

To conclude, we have discovered a correlation between the energy
shifts and site-energies in the spin-glass dynamics, and shown that
such a correlation can be sufficient for the dynamics to converge to a
metastable state in a large but finite system. Since in the
population-dynamics approach the converged state will have a nearly
continous $p(\lambda)$, while the typical metastable states have a
discontinous one, the success of a population-dynamics approach
implies the failure of the flat-measure one. Such success is only
possible because the population-dynamics converges: otherwise spins
would flip many times, and the Markov approximation would fail.

These considerations suggest an unusual picture of the origins of slow
dynamics in some complex systems. Disorder and frustration do play a role,
captured by the diffusion term, in preventing a fast convergence of the 
dynamics, but this role is limited by the drift. This causes the dynamics 
to converge long before it has time to thoroughly explore the state space, 
and so the Edwards hypothesis fails.

The Edwards hypothesis was shown to correctly predict the form of the
distribution of steady-state energies in simulations of tapping the SK
model in Ref.\ \onlinecite{dean2003}. Given our results, this agreement 
now poses an intriguing problem. Perhaps the true dynamical entropy
$\Sigma_{dyn}(E)$ has a similar energy dependence to that of the
flat-measure entropy $\Sigma_{edw}(E)$, so that the energy
distributions in tapping take similar forms. Since the states are very
different, however, it is unclear why this should occur. 

\begin{acknowledgments}
  We thank David Dean and David Sherrington for useful
  discussions. PRE acknowledges support from Sidney Sussex College,
  Cambridge and the Aspen Center for Physics, and RAB from the Royal
  Society of Edinburgh.

\end{acknowledgments}


\end{document}